\documentclass[aps,pra,onecolumn,superscriptaddress]{revtex4-2}
\usepackage{amssymb,amsmath,amsthm,mathrsfs,amsfonts,dsfont}
\usepackage{bm}
\usepackage{graphicx}
\usepackage{hyperref}
\usepackage{times}
\usepackage{subfigure}
\usepackage{hyperref} 
\usepackage{ulem}
\usepackage{color}
\hypersetup{colorlinks=true,citecolor=magenta,linkcolor=magenta,urlcolor=magenta}

\begin{document}


\title{Quantum metrology based on symmetry-protected adiabatic transformation:\\ Imperfection, finite time duration, and dephasing}


\author{Takuya Hatomura}
\email[]{takuya.hatomura.ub@hco.ntt.co.jp}
\affiliation{NTT Basic Research Laboratories \& NTT Research Center for Theoretical Quantum Physics, NTT Corporation, Kanagawa 243-0198, Japan}

\author{Atsuki Yoshinaga}
\email[]{a-yoshinaga@aist.go.jp}
\affiliation{Department of Physics, University of Tokyo, Chiba 277-8574, Japan}
\affiliation{National Institute of Advanced Industrial Science and Technology, Ibaraki 305-8568, Japan}

\author{Yuichiro Matsuzaki}
\email[]{matsuzaki.yuichiro@aist.go.jp}
\affiliation{National Institute of Advanced Industrial Science and Technology, Ibaraki 305-8568, Japan}

\author{Mamiko Tatsuta}
\email[]{mamiko.tatsuta@aist.go.jp}
\affiliation{National Institute of Advanced Industrial Science and Technology, Ibaraki 305-8568, Japan}


\date{\today}

\begin{abstract}
The aim of quantum metrology is to estimate target parameters as precisely as possible. 
In this paper, we consider quantum metrology based on symmetry-protected adiabatic transformation. 
We introduce a ferromagnetic Ising model with a transverse field as a probe and consider the estimation of a longitudinal field. 
Without the transverse field, the ground state of the probe is given by the Greenberger-Horne-Zeilinger state, and thus the Heisenberg limit estimation of the longitudinal field can be achieved through parity measurement. 
In our scheme, full information of the longitudinal field encoded on parity is exactly mapped to global magnetization by symmetry-protected adiabatic transformation, and thus the parity measurement can be replaced with global magnetization measurement. 
Moreover, this scheme requires neither accurate control of individual qubits nor that of interaction strength. 
We discuss the effects of the finite transverse field and nonadiabatic transitions as imperfection of adiabatic transformation. 
By taking into account finite time duration for state preparation, sensing, and readout, we also compare performance of the present scheme with a classical scheme in the absence and presence of dephasing. 
\end{abstract}

\pacs{}

\maketitle


%
%
\section{\label{Sec.intro}Introduction}

Precise estimation of parameters is desired for realizing upcoming quantum technologies such as quantum information processing. 
Quantum metrology is a promising method that offers higher precision sensing of target parameters than classical counterparts by exploiting entanglement~\cite{Toth2014,Degen2017,Pezze2018}. 
Appropriate entanglement among probe qubits enhances sensitivity, surpassing the standard quantum limit (SQL)~\cite{Caves1981,Giovannetti2004,Giovannetti2006}, which is known as the limit of classical sensors composed of independent qubits. 
In particular, the Greenberger-Horne-Zeilinger (GHZ) state~\cite{Greenberger1990,Mermin1990} achieves the ultimate precision called the Heisenberg limit in the absence of noise~\cite{Bollinger1996,Leibfried2004}.
Even under specific noise, the GHZ state can still beat the SQL~\cite{Matsuzaki2011,Chin2012,Chaves2013,Dur2014,Macieszczak2015,Zhou2018,Matsuzaki2018}. 
Considerable effort has been devoted to the development of entanglement generation and interferometry for practical use. 
However, application of entanglement-enhanced sensing is still limited due to the following reasons.

One of the major challenges in entanglement-enhanced sensing is to develop robust schemes against experimental imperfection. 
Typically, entanglement is created by gate operations~\cite{Leibfried2005,Neumann2008,Jones2009,DiCarlo2010,Neeley2010,Barends2014,Wei2020} or nonlinear interactions~\cite{Kitagawa1993,Agarwal1997,Molmer1999,Chumakov1999,Micheli2003,Pezze2009,Song2017,Song2019}. 
Controlled pulse sequences are required for adequately turning on/off gates or interactions to complete entanglement generation and to proceed to interferometry. It implies that complicated and precise setups are necessary in experiments.
Desirable schemes should not require accurate control of (individual) qubits.

In a ferromagnetic Ising model with a transverse field, macroscopic entanglement can be created in the ground state by adiabatically decreasing the transverse field~\cite{Cirac1998,Lee2006,Yukawa2018}. 
This process does not require accurate control of qubits. 
Moreover, this process is protected by symmetry, i.e., nonadiabatic transitions from even-parity energy eigenstates to odd-parity energy eigenstates do not take place because of parity conservation due to spin-flip symmetry~\cite{Xing2016,Hatomura2019,Hatomura2019a,Zhuang2020}. 
This suppression of nonadiabatic transitions protects the macroscopic entanglement from spontaneous symmetry breaking.

To use the macroscopic entanglement in the ferromagnetic Ising model for quantum metrology, parity measurement is required to extract information of a target parameter. 
Several ways to perform parity measurement exist. 
For example, we can obtain information of parity by post-processing data of single-qubit measurement on each qubit. 
However, operators to be measured in single-qubit measurement do not commute with the Hamiltonian (interaction term). 
In general, measurement of operators that do not commute with a given Hamiltonian is experimentally hard~\cite{Endo2020}, and thus we cannot perform single-qubit measurement unless interactions are turned off.

Recently, adiabatic transformation has been discussed~\cite{Dunningham2002,Huang2018,Haine2018} as a method of interaction-based readouts~\cite{Davis2016,Macri2016,Nolan2017,Yoshinaga2021} to change readout protocols. 
In particular, adiabatic transformation of the transverse field was introduced for the ferromagnetic Ising model to replace parity measurement with global magnetization measurement~\cite{Huang2018}. 
However, to achieve the Heisenberg limit, complicated optimization of the transverse field is necessary to adjust a redundant relative phase, which may not be suitable for practical use. 
Moreover, the dynamical range is limited, i.e., the Heisenberg limit scaling is achieved only for specific values of a target parameter.
It is also unclear for protocols based on adiabatic transformation whether or not they can beat the SQL when we take into account time duration for state preparation and readout.

In this paper, we consider a scheme for quantum metrology, in which we use the macroscopically entangled state in the ferromagnetic Ising model. 
This state is prepared by adiabatically decreasing the transverse field.
After exposing the macroscopically entangled state to a target longitudinal field, we adiabatically induce the transverse field again. 
This process is also protected by the symmetry, conserving the parity.
Consequently, we can extract full information of the parity by global magnetization measurement. 
For the strong transverse field, an operator to be measured commutes with the dominant part of the Hamiltonian (transverse field term), and thus our scheme is feasible in experiments. 
We discuss the effects of the finite transverse field and nonadiabatic transitions as imperfection of adiabatic transformation. 
By taking into account finite time duration for state preparation, sensing, and readout, we also compare performance of the present scheme with a classical scheme in the absence and presence of dephasing.

%
%
\section{\label{Sec.theory}Background}

%
%
\subsection{Quantum metrology}

In this section, we briefly review theory of quantum metrology (for details, see, Ref.~\cite{Toth2014,Degen2017,Pezze2018} and references therein). 
A typical procedure of quantum metrology is as follows. 
We prepare a probe state $|\Psi\rangle$ and expose it to a target parameter $\theta$ as $|\Psi_\theta\rangle=\exp(i\theta\hat{J})|\Psi\rangle$, where we assume that the generator $\hat{J}$ is the summation of $N$ local operators and its maximum (minimum) eigenvalue is $N/2$ ($-N/2$). 
Then, we measure an observable of the probe $\hat{A}$ and obtain a measurement outcome. 
By repeating this process many times, we estimate the target parameter $\theta$. 
The uncertainty of the estimation is given by the error-propagation formula
\begin{equation}
\delta\theta_\mathrm{est}=\frac{(\Delta\hat{A})_\theta}{\left|\frac{\partial\langle\hat{A}\rangle_\theta}{\partial\theta}\right|\sqrt{M}},
\end{equation}
where $(\Delta\hat{A})_\theta=\sqrt{\langle\hat{A}^2\rangle_\theta-\langle\hat{A}\rangle_\theta^2}$ and $\langle\hat{A}\rangle_\theta=\langle\Psi_\theta|\hat{A}|\Psi_\theta\rangle$. 
Here, $M$ is the number of measurement. 
According to the Cram\'er-Rao bound, the uncertainty of the estimation is lower bounded by the quantum Fisher information, i.e., $\delta\theta_\mathrm{est}\ge1/\sqrt{MF_Q}$, where $F_Q=4\langle\partial_\theta\Psi_\theta|(1-|\Psi_\theta\rangle\langle\Psi_\theta|)|\partial_\theta\Psi_\theta\rangle$ is the quantum Fisher information. 
For a probe state satisfying $\langle\Psi|\hat{J}^2|\Psi\rangle=N^2/4$ 
and $\langle\Psi|\hat{J}|\Psi\rangle=0$, the Cram\'er-Rao bound provides the ultimate limit $\delta\theta_\mathrm{HL}=1/N\sqrt{M}$, which is the Heisenberg limit. 
Note that the Cram\'er-Rao bound also provides the limit of classical sensors composed of separable states $\delta\theta_\mathrm{SQL}=1/\sqrt{NM}$, which is the SQL.

In physical setups, the target parameter $\theta$ is the product of a physical target parameter $\omega$ and time duration for sensing (interaction with the target parameter) $T_\mathrm{int}$, i.e., $\theta=\omega T_\mathrm{int}$. 
Then, the uncertainty of the estimation is given by
\begin{equation}
\delta\omega_\mathrm{est}=\frac{(\Delta\hat{A})_{\theta=\omega T_\mathrm{int}}}{\left|\frac{\partial\langle\hat{A}\rangle_{\theta=\omega T_\mathrm{int}}}{\partial\omega}\right|\sqrt{M}}. 
\label{Eq.uncertainty}
\end{equation}
The Heisenberg limit and the SQL are also rewritten as
\begin{equation}
\delta\omega_\mathrm{HL}=\frac{1}{N\sqrt{M}T_\mathrm{int}}
\label{hl}
\end{equation}
and
\begin{equation}
\delta\omega_\mathrm{SQL}=\frac{1}{\sqrt{NM}T_\mathrm{int}},
\label{sql}
\end{equation}
respectively. 
Moreover, for a given total time $T$, the number of measurement can be expressed in terms of time duration for state preparation $T_\mathrm{prep}$, sensing $T_\mathrm{int}$, and readout $T_\mathrm{read}$, as
\begin{equation}
M=\frac{T}{T_\mathrm{prep}+T_\mathrm{int}+T_\mathrm{read}}. 
\end{equation}
When $T_\mathrm{prep}\to0$, $T_\mathrm{read}\to0$, and $T_\mathrm{int}\to T$, these limits are minimized as
\begin{equation}
\delta\omega_\mathrm{HL,min}=\frac{1}{NT},
\label{Eq.HL.min}
\end{equation}
and
\begin{equation}
\delta\omega_\mathrm{SQL,min}=\frac{1}{\sqrt{N}T}.
\label{Eq.SQL.min}
\end{equation}
Note that the minimized Heisenberg limit (\ref{Eq.HL.min}) is not realistic because a statistical average for obtaining the expectation value of the observable is neglected. 
Therefore, another minimized Heisenberg limit
\begin{equation}
\delta\omega_\mathrm{HL,min}^\ast=\frac{1}{N\sqrt{T_\mathrm{int}T}},
\label{Eq.HL.min2}
\end{equation}
is also used, where $T_\mathrm{prep}\to0$ and $T_\mathrm{read}\to0$, but $T_\mathrm{int}\ll T$ so that $M\gg1$.

%
%
\subsection{Model}

As a probe system, we consider the following infinite-range Ising model with a transverse field
\begin{equation}
\hat{\mathcal{H}}=-\frac{1}{2}J\sum_{i,j=1}^N\hat{Z}_i\hat{Z}_j-h^x\sum_{i=1}^N\hat{X}_i,
\label{Eq.ham}
\end{equation}
where we express Pauli matrices as $\{\hat{X},\hat{Y},\hat{Z}\}$, and $J$ and $h^x$ are the interaction strength and the amplitude of the transverse field, respectively. 
We assume that $h^x$ is tunable, while $J$ is fixed. 
This is a reasonable assumption for many physical systems. 
In addition, we assume $N$ to be even for simplicity. 
Our purpose is to estimate a target longitudinal field $h^z$. 
In a sensing process, 
\begin{equation}
\hat{V}=-h^z\sum_{i=1}^N\hat{Z}_i 
\end{equation}
is added to the Hamiltonian (\ref{Eq.ham}). 
Here, the relationship between the target parameter $\omega$ in the previous section and the target longitudinal field $h^z$ is given by $\omega=2h^z$.

For convenience, we use eigenvectors 
\begin{equation}
\hat{S}_W|N/2,m\rangle_W=m|N/2,m\rangle_W\quad(W=X,Y,Z)
\end{equation}
of collective spin operators
\begin{equation}
\hat{S}_W=\frac{1}{2}\sum_{i=1}^N\hat{W}_i\quad(W=X,Y,Z)
\label{Eq.collective}
\end{equation}
to express energy eigenstates of the Hamiltonian (\ref{Eq.ham}).
Here we suppose that the system is confined in the maximum spin subspace satisfying $\sum_{W=X,Y,Z}\hat{S}_W^2=N/2\times(N/2+1)$, i.e., $m=-N/2,-N/2+1,\dots,N/2$.

This system (\ref{Eq.ham}) conserves the parity
\begin{equation}
\hat{\Pi}=\prod_{i=1}^N\hat{X}_i,
\label{Eq.parity}
\end{equation}
i.e., the commutation relation between the Hamiltonian (\ref{Eq.ham}) and the parity operator (\ref{Eq.parity}) becomes zero. That is,
\begin{equation}
[\hat{\mathcal{H}},\hat{\Pi}]=0 
\label{Eq.parity.conserve}
\end{equation}
for any $h^x$ (see, e.g., Ref.~\cite{Xing2016,Hatomura2019,Hatomura2019a,Zhuang2020}). 
Therefore, $(N+1)$ energy eigenstates of the Hamiltonian (\ref{Eq.ham}) in the maximum spin subspace are classified into two sets, $\{|\psi_n(h^x)\rangle\}_{n=0}^{N/2}$ with the parity $\hat{\Pi}=+1$ and $\{|\phi_n(h^x)\rangle\}_{n=0}^{N/2-1}$ with the parity $\hat{\Pi}=-1$, in the ascending order of energy, respectively. 
These energy eigenstates are given by
\begin{equation}
\left\{
\begin{aligned}
&|\psi_n(\infty)\rangle=|N/2,N/2-2n\rangle_X, \\
&|\phi_n(\infty)\rangle=|N/2,N/2-(2n+1)\rangle_X 
\end{aligned}
\right.
\label{Eq.eigen.infty}
\end{equation}
in the $h^x\to\infty$ limit and
\begin{equation}
\left\{
\begin{aligned}
&|\psi_n(0)\rangle=\frac{1}{\sqrt{2}}(|N/2,N/2-n\rangle_Z+|N/2,n-N/2\rangle_Z), \\
&|\phi_n(0)\rangle=\frac{1}{\sqrt{2}}(|N/2,N/2-n\rangle_Z-|N/2,n-N/2\rangle_Z)
\end{aligned}
\right.
\label{Eq.eigen.zero}
\end{equation}
for $n=0,1,\dots,N/2-1$ and $|\psi_{N/2}(0)\rangle=|N/2,0\rangle_Z$ in the $h^x\to0$ limit. 
Notably, the degenerate ground states $|\psi_0(0)\rangle$ and $|\phi_0(0)\rangle$, which are known as the GHZ states, can achieve the Heisenberg limit (\ref{hl}) by parity measurement~\cite{Bollinger1996}.

For example, we can obtain the expectation value of the parity (\ref{Eq.parity}) by implementing single-qubit measurement of $\hat{X}$ on each qubit and multiplying the measurement outcomes, and by averaging it for many independent and identically distributed samples. 
However, single-qubit measurement of $\hat{X}$ is nontrivial for the present model because each $\hat{X}$ does not commute with the interaction term of the Hamiltonian. 
If the interaction term is much smaller than the resonant frequency of qubits, we can perform single-qubit rotation along the $y$-axis by $\pi/2$ and subsequent single-qubit measurement of $\hat{Z}$, which commutes with the interaction term of the Hamiltonian, for each qubit. The measurement outcome is equivalent to $\hat{X}$ of the original state. 
However, when the interaction term is as large as or larger than the resonant frequency of qubits, we cannot use this method. 
Other approaches are necessary to measure the parity.

%
%
\subsection{\label{Sec.method.ideal}State preparation and readout based on symmetry-protected adiabatic transformation}

In this section, we explain our scheme with a reasonable readout protocol extracting full information of the parity. 
First, we generate $|\psi_0(0)\rangle$ by adiabatic transformation, i.e., we prepare the trivial ground state $|\psi_0(\infty)\rangle$ as the initial state and adiabatically change the transverse field $h^x$ from infinity to zero~\cite{Cirac1998,Lee2006,Yukawa2018}. 
We then expose the system to the target longitudinal field $h^z$ during a time interval $T_\mathrm{int}$. 
As mentioned in the previous section, we do not assume a situation where the interaction term can be turned off during sensing. 
Finally, we adiabatically change the transverse field $h^x$ again to infinity, and then the probe state becomes
\begin{equation}
|\Psi_{\theta=2h^zT_\mathrm{int}}\rangle=\cos(h^zNT_\mathrm{int})|\psi_0(\infty)\rangle+\sin(h^zNT_\mathrm{int})e^{i\alpha}|\phi_0(\infty)\rangle 
\label{Eq.readoutstate}
\end{equation}
except for a global phase factor~\cite{Huang2018}. 
Here, $\alpha$ is a relative phase accompanying the adiabatic transformation of the transverse field $h^x$. 
In Ref.~\cite{Huang2018}, global magnetization measurement of $\hat{S}_Z$ was discussed, but we consider global magnetization measurement of $\hat{S}_X$ (or, projection measurement of $\hat{S}_X=N/2$, i.e., measuring $P=|\langle\Psi_{\theta=2h^zT_\mathrm{int}}|N/2,N/2\rangle_X|^2$). 
The uncertainty of the estimation achieves the Heisenberg limit (\ref{hl}), 
\begin{equation}
\delta h^z_\mathrm{est}=\frac{1}{2N\sqrt{M}T_\mathrm{int}}. 
\end{equation}

Here we explain key points of the present scheme. 
The first point is that adiabatic transformation for state preparation and readout is protected by symmetry. 
That is, owing to the spin-flip symmetry, nonadiabatic transitions between the ground state and the first excited state (the degenerate ground state for small $h^x$) do not take place~\cite{Xing2016,Hatomura2019,Hatomura2019a,Zhuang2020}. 
It mitigates the adiabatic condition. 
The second point is that the full information of the target longitudinal field $h^z$, which is encoded on the amplitude of different parity eigenstates with a factor $N$ in the sensing process, is completely mapped to the amplitude of different magnetization eigenstates of $\hat{S}_X$ because of the parity conservation due to the spin-flip symmetry. 
Therefore, the present scheme achieves the Heisenberg limit. 
It is also an important point that the observable $\hat{S}_X$ commutes with the dominant part of the Hamiltonian.

Global magnetization measurement of $\hat{S}_Z$ discussed in Ref.~\cite{Huang2018} leads to similar results, but complicated nonlinear adjustment of the relative phase $\alpha$ is required and the dynamical range is limited (see Appendix~\ref{Sec.meas.Sz}). 
Note that if we can apply the $\pi/2$ pulse along the $y$-axis, we can replace $\hat{S}_X$ measurement with $\hat{S}_Z$ measurement.

%
%
\subsection{\label{Sec.noise}Phase shift}

While the present scheme achieves the Heisenberg limit for any $h^z$, both the denominator and the numerator in Eq.~(\ref{Eq.uncertainty}) vanish for $h^z\ll1$ because the expectation value of $\hat{S}_X$ is the sine-squared function (the expectation value of projection measurement $P$ is the cosine-squared function). 
However, in noisy situations, the numerator typically has a finite value, while the denominator is infinitesimal for small $h^z$, resulting in divergence of the uncertainty. 
For example, the numerator becomes large when readout measurement becomes noisy~\cite{Taylor2008,Kitazawa2017}.

To avoid such a problem, we introduce a phase shift. 
The target parameter can be divided into two parts, $h^z=h^z_k+h^z_u$, where $h^z_k$ is a known part and $h^z_u$ is an unknown part. 
By performing prior estimation with a classical sensor, we can assume that an approximate value of $h^z$ is known, i.e., $h^z_k\approx h^z$ and $h^z_u\ll 1$. 
Then, we try to estimate $h^z_u$ by entanglement-enhanced sensing for further improvement of precision. 
As a phase shift, we add an offset $h^z_0$ so that $2(h^z_k+h_0^z)NT_\mathrm{int}=(2n+1)\pi/2$ with an integer $n$. 
Then, the denominator of Eq.~(\ref{Eq.uncertainty}) turns into $|\partial\langle\hat{S}_X\rangle_{\theta=2h^zT_\mathrm{int}}/\partial h^z|=NT_\mathrm{int}|\cos(2h^z_uNT_\mathrm{int})|$ for global magnetization measurement ($|\partial P/\partial h^z|=NT_\mathrm{int}|\cos(2h^z_uNT_\mathrm{int})|$ for projection measurement), which does not vanish for small $h^z_u$. 
This phase shift is necessary for beating the SQL when we take into account finite transverse field, dephasing, and nonadiabatic transitions.

%
%
\subsection{\label{Sec.deph.back}Dephasing}

Dephasing during the sensing process is a main obstacle for quantum-enhanced sensing. 
Here we explain the effect of time-inhomogeneous dephasing (non-Markovian dephasing) during the sensing process. 
Note that, in the following discussion, we always apply the phase shift discussed in the previous section.

As a reference scheme, we consider an ensemble of $N$ qubits without entanglement and assume that time duration for state preparation and readout is negligibly small. 
In the presence of non-Markovian dephasing, the uncertainty of the estimation is given by $\delta h^z_\mathrm{est}=e^{\Gamma^2T_\mathrm{int}^2/2}/2\sqrt{NT_\mathrm{int}T}$, where $\Gamma$ is the dephasing rate (the decay rate of the off-diagonal elements). 
This uncertainty of the estimation is minimized when $T_\mathrm{int}^2=1/2\Gamma^2$, and then the reference scheme gives the minimized SQL under dephasing
\begin{equation}
\delta h^z_\mathrm{SQL,deph,min}=\frac{(2e\Gamma^2)^{1/4}}{2\sqrt{NT}}.
\end{equation}
In schemes using the GHZ state, we also take into account time-inhomogeneous dephasing during the sensing process, and then the uncertainty of the estimation is given by
\begin{equation}
\delta h^z_\mathrm{est}=\frac{\sqrt{T_\mathrm{prep}+T_\mathrm{int}+T_\mathrm{read}}e^{\Gamma^2NT_\mathrm{int}^2/2}}{2NT_\mathrm{int}\sqrt{T}}. 
\end{equation}
When time duration for state preparation and readout is negligibly small, it is minimized for $T_\mathrm{int}^2=1/2\Gamma^2N$, and then we obtain the Zeno limit scaling $\delta h^z_\mathrm{est}=(2e\Gamma^2)^{1/4}/2N^{3/4}\sqrt{T}$~\cite{Matsuzaki2011,Chin2012}.
Moreover, with this sensing time, we can still beat the SQL in the sense of scaling when time duration for state preparation and readout is $T_\mathrm{prep}+T_\mathrm{read}<\mathcal{O}(N^0)$~\cite{Dooley2016a} although such fast state preparation and readout may not be realistic for many-body entanglement creation. 
In the entanglement scheme with $T_\mathrm{prep}+T_\mathrm{read}\ge\mathcal{O}(N^0)$, constant factor improvement over the SQL is possible when the equality is satisfied, and conditional improvement over the SQL is still possible when the number of qubits $N$ is smaller than a certain threshold~\cite{Dooley2016a}. 
Even in this case, by preparing sub-ensembles consisting of $N^\prime$ ($<N$) qubits, where the number of qubits in each sub-ensemble $N^\prime$ satisfies the threshold, we can perform entanglement-enahanced sensing with large number of qubits $N$~\cite{Dooley2016a}.

%
%
\section{Results}

%
%
\subsection{\label{Sec.real}Finite transverse field}

We considered the infinite transverse field in Sec.~\ref{Sec.method.ideal}. 
In this section, we discuss the case of a finite transverse field, i.e., we change the transverse field $h^x$ from $h^x_0$ (0) to $0$ ($h^x_0$) in the state preparation (readout) process. 
In particular, we derive conditions of the transverse field for achieving the Heisenberg limit scaling.

Let us discuss two approaches to prepare the initial state. 
The first approach is as follows: for $h^x_0/JN\gg1$, we prepare the ground state $|\psi_0(h^x_0)\rangle$ as the initial state, which can be done by cooling the system because of large energy gap. 
However, in this case, long operation time is required to adiabatically change the transverse field from large $h^x_0$ to 0 and from 0 to large $h^x_0$.
The other approach is as follows: we apply a strong magnetic field $h^x/JN\gg1$ and perform projection measurement of $|\psi_0(\infty)\rangle=|N/2,N/2\rangle_X$, and implement sudden quench to $h^x_0$ satisfying $h^x_0/JN\approx1$. 
In this case, the operation time to satisfy the adiabatic condition can be shorter than the first approach, while the initial state becomes $|\psi_0(\infty)\rangle$. 
This state is not the ground state of the given Hamiltonian, but close to it as discussed later. 
Note that $h^x/JN=1$ is the critical point in the thermodynamic limit, and thus we cannot prepare the ground state by cooling because of small energy gap.

Similarly to Sec.~\ref{Sec.method.ideal}, we adiabatically turn off the transverse field from $h^x_0$ to 0, expose the system to the target longitudinal field $h^z$, and adiabatically turn on the transverse field from $0$ to $h^x_0$. 
The probe state becomes
\begin{equation}
|\Psi_{\theta=h^zT_\mathrm{int}}\rangle=\cos(h^zNT_\mathrm{int})|\psi_0(h^x_0)\rangle+\sin(h^zNT_\mathrm{int})e^{i\alpha^\prime}|\phi_0(h^x_0)\rangle,
\end{equation}
in the former case of initial state preparation, and,
\begin{equation}
|\Psi_{\theta=h^zT_\mathrm{int}}\rangle=\sum_{n=0}^{N/2}g_ne^{i\gamma_n}\{\cos[h^z(N-2n)T_\mathrm{int}]|\psi_n(h^x_0)\rangle+\sin[h^z(N-2n)T_\mathrm{int}]e^{i\alpha_n}|\phi_n(h^x_0)\rangle\},
\end{equation}
in the latter case, where $g_n=\langle\psi_n(h^x_0)|\psi_0(\infty)\rangle=\langle\psi_n(h^x_0)|N/2,N/2\rangle_X$ is the overlap between the initial state and the ground state. 
Here, $\alpha^\prime$, $\alpha_n$, and $\gamma_n$ are relative phases.

Finally we perform the projection measurement of $|\psi_0(\infty)\rangle=|N/2,N/2\rangle_X$ and obtain
\begin{equation}
P=|g_0|^2\cos^2(h^zNT_\mathrm{int}),
\end{equation}
in the former case, and,
\begin{equation}
P=\left|\sum_{n=0}^{N/2}|g_n|^2e^{i\gamma_n}\cos[h^z(N-2n)T_\mathrm{int}]\right|^2 
\label{Eq.projection.realistic}
\end{equation}
in the latter case, as the survival probability of this measurement.

\begin{figure}
\includegraphics[width=8cm]{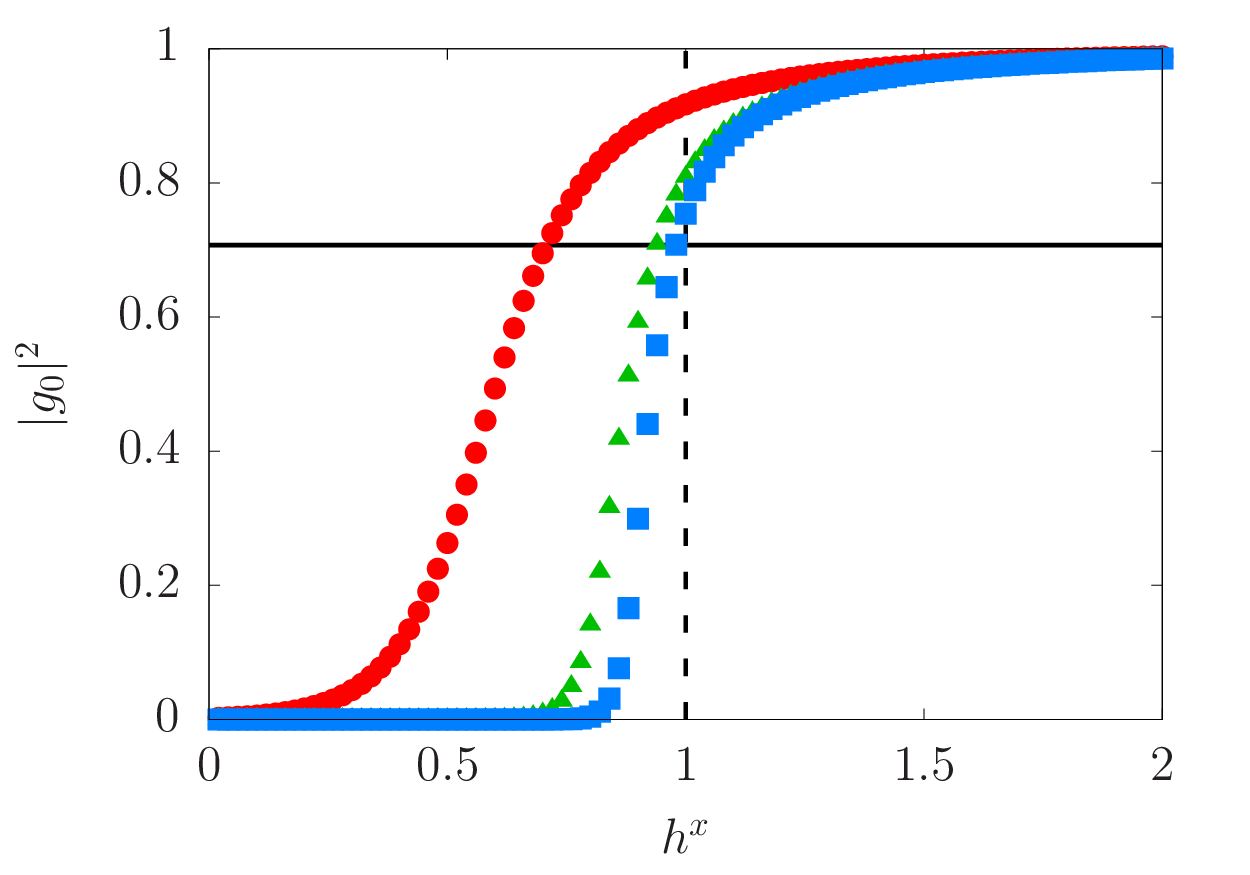}
\caption{\label{Fig.overlap}Overlap between the initial state and the ground state $|g_0|^2=|\langle\psi_0(h^x)|\psi_0(\infty)\rangle|^2$. The horizontal axis is the transverse field $h^x$ in units of $JN$. Here, (red circles) $N=10$, (green triangles) $N=50$, and (blue squares) $N=100$. The solid horizontal line represents the threshold $|g_0|^4=1/2$ and the dashed vertical line represents the critical point. }
\end{figure}

We can immediately find an upper bound for the uncertainty of the estimation (\ref{Eq.uncertainty}), 
\begin{equation}
\delta h^z_\mathrm{est}\le\frac{1}{2N\sqrt{M}T_\mathrm{int}|g_0|^2|\sin(2h^zNT_\mathrm{int})|},
\label{Eq.bound1}
\end{equation}
in the former case, and after some calculations, we can also derive an upper bound
\begin{equation}
\delta h^z_\mathrm{est}\le\frac{1}{2N\sqrt{M}T_\mathrm{int}(2|g_0|^4-1)\sin(2h^zNT_\mathrm{int})} 
\label{Eq.uncertainty.bound}
\end{equation}
in the latter case, for $|g_0|^4>1/2$ when the condition $0\le 2h^zNT_\mathrm{int}\le\pi/2$ is satisfied (see Appendix~\ref{Sec.bound.derivation} for derivation). 
Notably, the factor $\sin(2h^zNT_\mathrm{int})$ becomes unity when we consider the phase shift discussed in Sec.~\ref{Sec.noise} and the right-hand sides of Eqs.~(\ref{Eq.bound1}) and (\ref{Eq.uncertainty.bound}) exactly coincide with the Heisenberg limit when $|g_0|^2=1$. 
These bounds guarantee the Heisenberg limit scaling when the overlap between the initial state and the ground state, $|g_0|^2=|\langle\psi_0(h_0^x)|\psi_0(\infty)\rangle|^2$, satisfies $|g_0|^2=\Theta(N^0)$ in the former case and $2|g_0|^4-1=\Theta(N^0)$ in the latter case, respectively. 
We plot the overlap $|g_0|^2$ and the latter threshold $|g_0|^4=1/2$ in Fig.~\ref{Fig.overlap}. 
We find that the initial condition $h^x_0/JN\approx2$ is large enough for achieving the Heisenberg limit scaling, and the condition $h_0^x/JN=1$ is enough for beating the SQL when $N\le100$.

%
%
\subsection{Finite time duration for state preparation and readout}

In this section, we take into account finite time duration for state preparation and readout, and discuss conditions for beating the minimized SQL (\ref{Eq.SQL.min}) and for achieving similar scaling to the minimized Heisenberg limits (\ref{Eq.HL.min}) and (\ref{Eq.HL.min2}). 
From the derivation of these minimized limits, faster implementation of state preparation and readout than that of sensing seems necessary, and then one may suspect that critical slowing down could spoil the effectiveness of our scheme as in the case of criticality-based quantum metrology~\cite{Rams2018,Gietka2021}, i.e., long time duration for state preparation and readout based on adiabatic transformation restricts the sensing time $T_\mathrm{int}$ and the total process may be beaten by even the SQL. 
However, our conclusion is that time duration for state preparation and readout is not necessarily shorter than that for sensing. 
For simplicity, we assume that $T_\mathrm{prep}=T_\mathrm{read}=T_a$ in the present paper.

In the present model, the energy gap appears at the critical point $h^x/JN=1$, and it scales as $\Delta E=\mathcal{O}(N^{-1/3})$~\cite{Botet1983,Caneva2008,Yoshimura2014}. 
Therefore, according to the adiabatic condition, time duration for state preparation and readout is roughly given by 
\begin{equation}
2T_a=T_\mathrm{prep}+T_\mathrm{read}=CN^{2/3}
\end{equation}
with an $N$-independent constant $C$ (see also, Appendix~\ref{Sec.ad.time}). 
To satisfy $T\gg T_\mathrm{int}+2T_a$, the total time $T$ must scale as at least 
\begin{equation}
T=\tilde{C}N^{2/3}
\end{equation}
with an $N$-independent constant $\tilde{C}\gg C$. 
Then we find that the condition for beating the SQL, i.e., $\delta h_\mathrm{est}^z<\delta h_\mathrm{SQL,min}^z$, is given by 
\begin{equation}
T_\mathrm{int}>\frac{T}{2N}+\sqrt{\frac{2T_aT}{N}+\left(\frac{T}{2N}\right)^2}=\sqrt{C\tilde{C}}N^{1/6}+\mathcal{O}(N^{-1/3}). 
\end{equation}
Therefore, even if the interaction time with the target field is much shorter than state preparation and readout, i.e., $T_\mathrm{int}<2T_a=\mathcal{O}(N^{2/3})$, we can beat the SQL by setting $T_\mathrm{int}>\mathcal{O}(N^{1/6})$ for $T=\mathcal{O}(N^{2/3})$.

Next, by increasing time duration for sensing, we show how the uncertainty is improved and when the Heisenberg limit scaling is achieved. 
To elucidate these points, we rewrite the uncertainty of the estimation as $\delta h^z_\mathrm{est}=\delta h^z_\mathrm{SQL,min}/\eta$ and $\delta h^z_\mathrm{est}=\delta h^{z\ast}_\mathrm{HL,min}/\eta^\prime$, where $\eta$ and $\eta^\prime$ are given by
\begin{equation}
\eta=\sqrt{\frac{NT_\mathrm{int}^2}{T(T_\mathrm{int}+2T_a)}}
\end{equation}
and
\begin{equation}
\eta^\prime=\sqrt{\frac{T_\mathrm{int}}{T_\mathrm{int}+2T_a}},
\end{equation}
respectively. 
We set the sensing time $T_\mathrm{int}$ as 
\begin{equation}
T_\mathrm{int}=\sqrt{C\tilde{C}}N^{1/6+\epsilon}, 
\end{equation}
where $\epsilon\ge0$. 
Since $T_{\mathrm{int}}\leq T=\mathcal{O}(N^{2/3})$, we must keep $\epsilon\leq 1/2$.  
Then, we find
\begin{equation}
\eta=\frac{N^\epsilon}{\left(1+\sqrt{\tilde{C}/C}N^{-1/2+\epsilon}\right)^{1/2}}=N^\epsilon+\mathcal{O}(N^{-1/2+2\epsilon})
\end{equation}
for $0<\epsilon<1/2$, and
\begin{equation}
\eta^\prime=\frac{1}{1+\sqrt{C/\tilde{C}}}\approx1
\end{equation}
for $\epsilon=1/2$. 
That is, we can beat the SQL and improve the uncertainty by $N^\epsilon$ for $0<\epsilon<1/2$ and achieve the Heisenberg limit scaling for $\epsilon=1/2$. 
We can also find similar results for Eq.~(\ref{Eq.HL.min}) (see, Appendix~\ref{Sec.finite.time.duration}).

%
%
\subsection{Dephasing}

As mentioned in the previous section, time duration for state preparation and readout in our scheme is given by $T_\mathrm{prep}+T_\mathrm{read}=CN^{2/3}$. 
Therefore, our scheme cannot achieve even the SQL scaling in the presence of dephasing as discussed in Sec.~\ref{Sec.deph.back}. 
However, our scheme can still beat the SQL for specific number of qubits $N$.

\begin{figure}
\includegraphics[width=8cm]{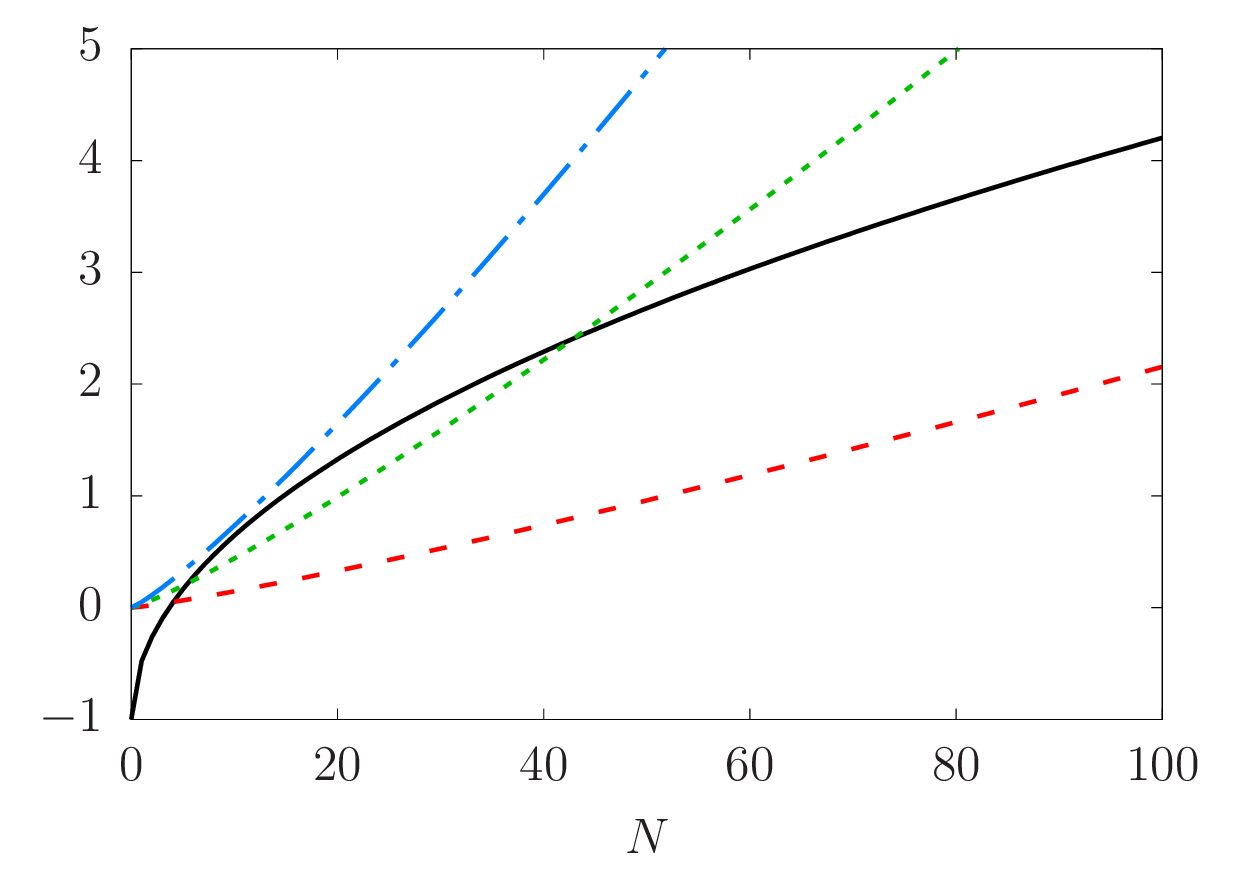}
\caption{\label{Fig.deph}Both sides of Eq.~(\ref{Eq.cond.SQL.deph}) against the system size $N$.
The black solid curve represents the right-hand side of it, and the red dashed curve, the green dotted curve, and the blue dash-dotted curve represent the left-hand side of it for $\Gamma C=0.01$, $0.03$, and $0.05$, respectively. }
\end{figure}

In the presence of dephasing, time duration for sensing must be much smaller than that for state preparation and readout, $T_\mathrm{int}\ll T_\mathrm{prep}+T_\mathrm{read}$, and thus the uncertainty of the estimation is roughly given by $\delta h^z_\mathrm{est}\approx\sqrt{T_\mathrm{prep}+T_\mathrm{read}}e^{\Gamma^2NT_\mathrm{int}^2/2}/2NT_\mathrm{int}\sqrt{T}$, which is minimized for $T_\mathrm{int}^2=1/\Gamma^2N$.
For this sensing time, the condition for beating the mimimized SQL, i.e., $\delta h^z_\mathrm{est}<\delta h^z_\mathrm{SQL,deph,min}$, is given by
\begin{equation}
\Gamma CN^{7/6}<\frac{\sqrt{2}}{e}N^{1/2}-1. 
\label{Eq.cond.SQL.deph}
\end{equation}
For various values of $\Gamma C$, we plot the left-hand side of this equation with the right-hand side in Fig.~\ref{Fig.deph}. 
We find that, for small $\Gamma C$, we can beat the SQL for specific number of qubits $N$.

Now, we discuss the value of $\Gamma C$. 
According to the analysis in Ref.~\cite{Caneva2008}, we find that the constant $C$ is given by $JNC=(h_0^x/JN)\bar{C}$ with a dimensionless constant $\bar{C}$ (see, Appendix~\ref{Sec.ad.time}). 
The dimensionless constant $\bar{C}$ is roughly given by $\mathcal{O}(1)$ or $\mathcal{O}(10)$ depending on the required fidelity. 
As we show in Sec.~\ref{Sec.real}, $h_0^x/JN=2$ is large enough for our scheme, and thus $JNC$ can also be $\mathcal{O}(1)$ or $\mathcal{O}(10)$. 
Therefore, for beating the SQL with dozen or several hundreds of qubits, it is expected that $\Gamma/JN$ should be $\mathcal{O}(10^{-2})$ or $\mathcal{O}(10^{-3})$ at worst.

%
%
\subsection{\label{Sec.nonad}Nonadiabatic time scale}

Finally, we discuss performance of our scheme with small system size $N=10,20,\dots,100$ in nonadiabatic time scale. 
We set $h_0^x=JN$ and change the transverse field $h^x$ as $h^x=h_0^x\cos(\pi t/2T_a)$ for $0\le t\le T_a$, which was introduced as coherent driving in Ref.~\cite{Yukawa2018} and is similar to a geometrically optimal schedule~\cite{Hatomura2019a}. 
Under this transverse field, we can shorten the operation time $T_a$ because nonadiabatic transitions and interference result in high fidelity to the GHZ state even in nonadiabatic time scale~\cite{Yukawa2018}. 
We also change the transverse field $h^x$ as $h^x=h_0^x\sin\{\pi[t-(T_a+T_\mathrm{int})]/2T_a\}$ for $T_a+T_\mathrm{int}\le t\le2T_a+T_\mathrm{int}$. 
 In the following numerical simulations, we set $JN=1$ and omit $M$.

\begin{figure}
\includegraphics[width=8cm]{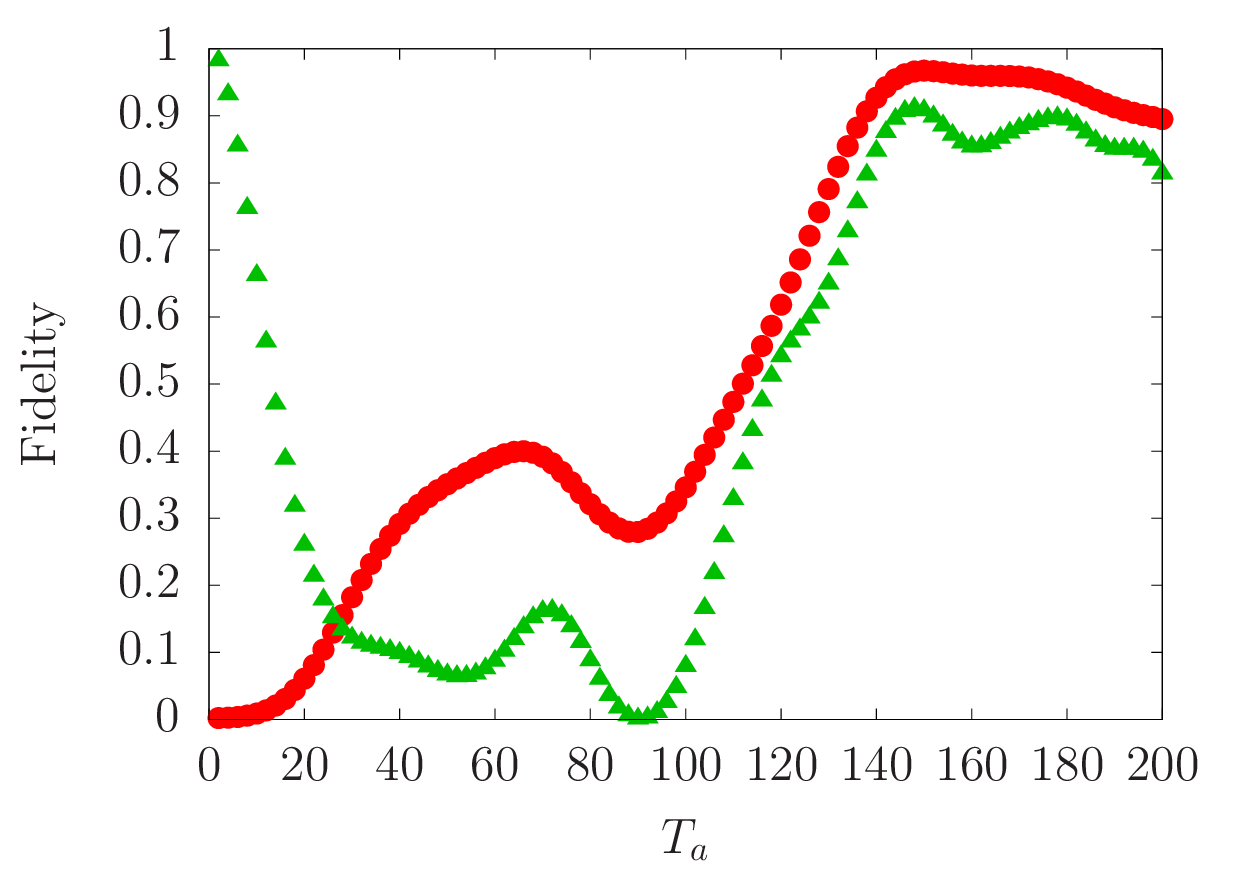}
\caption{\label{Fig.fidN10}Fidelity to the GHZ state at time $t=T_a$ (red circles) and to the initial state at time $t=2T_a+T_\mathrm{int}=2T_a$ (green triangles) for $N=10$. The horizontal axis is the operation time $T_a$ in units of $(2JN^2)^{-1}$. }
\end{figure}

First, we optimize the operation time $T_a$ for the sensing time $T_\mathrm{int}=0$. 
We plot (red circles) the fidelity of the probe state to the GHZ state $|\psi_0(0)\rangle$ at the time $t=T_a$ and (green triangles) that to the initial state $|\psi_0(\infty)\rangle$ at the time $t=2T_a+T_\mathrm{int}=2T_a$ for $N=10$ in Fig.~\ref{Fig.fidN10}. 
Here, interference appears when nonadiabatic transitions take place, and thus these quantities show oscillating behavior. 
We find a locally optimal operation time $T_a\approx150(2JN^2)^{-1}$ showing high fidelity to the GHZ state ($\sim0.97$) and to the initial state ($\sim0.91$).

\begin{figure}
\includegraphics[width=8cm]{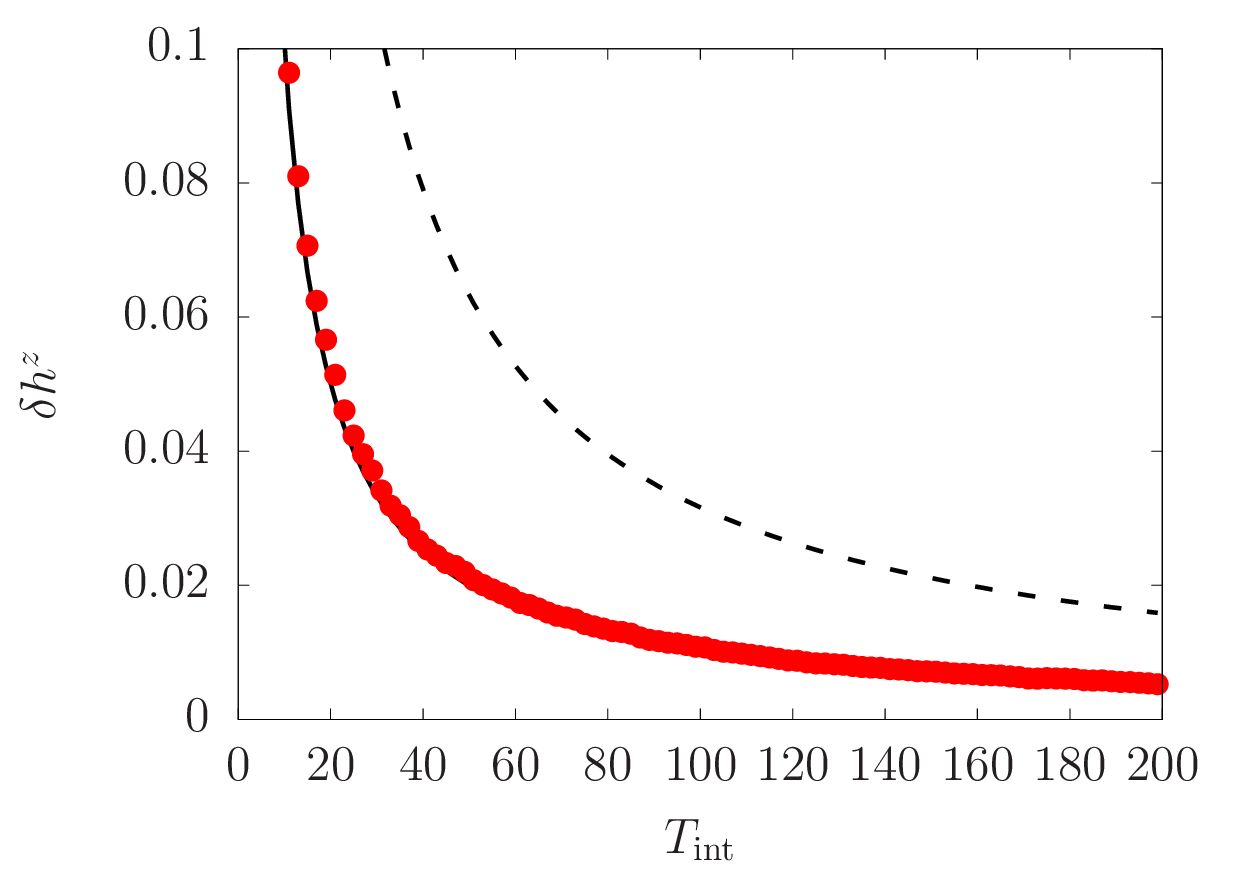}
\caption{\label{Fig.uncertaintyN10} Uncertainty of the estimation for $N=10$. The horizontal axis is the sensing time $T_\mathrm{int}$ in units of $(2JN^2)^{-1}$ and the vertical axis is the uncertainty of estimation $\delta h^z_\mathrm{est}$ in units of $JN$. The solid and dashed curves represent the Heisenberg limit, $\delta h^z_\mathrm{HL}$, and the SQL, $\delta h^z_\mathrm{SQL}$, respectively.}
\end{figure}

Now, we set $T_a=150(2JN^2)^{-1}$ and study the uncertainty of the estimation (\ref{Eq.uncertainty}) for the infinitesimal target parameter $h^z_u$ with the phase shift discussed in Sec.~\ref{Sec.noise}. 
We calculate the denominator of Eq.~(\ref{Eq.uncertainty}) by finite difference, $\partial P/\partial h^z_u\approx(P|_{h^z_u=10^{-10}}-P|_{h^z_u=0})/10^{-10}$. 
The sensing time $T_\mathrm{int}$ contributes to relative phases between different levels, and it affects the uncertainty of the estimation (\ref{Eq.uncertainty}) [see Eq.~(\ref{Eq.projection.realistic})]. 
Therefore, we plot the uncertainty of the estimation (\ref{Eq.uncertainty}) with respect to $T_\mathrm{int}$ in Fig.~\ref{Fig.uncertaintyN10}. 
We find that the uncertainty is very close to the Heisenberg limit. 
Indeed, the uncertainty of the estimation (\ref{Eq.uncertainty}) achieves $\delta h^z_\mathrm{est}\approx1.07/2NT_\mathrm{int}$ on average for $(2JN^2)T_\mathrm{int}=1,3,5,\dots,199$. 
Here, $(h_k^z+h_0^z)/JN=\pi/2$. 
Note that $1.07\approx(0.93)^{-1}$ and thus it is smaller than that expected from the fidelity to the GHZ state ($\sim0.97$) and a little bit larger than that expected from the fidelity to the initial state ($\sim0.91$) for $T_\mathrm{int}=0$.

\begin{figure}
\includegraphics[width=8cm]{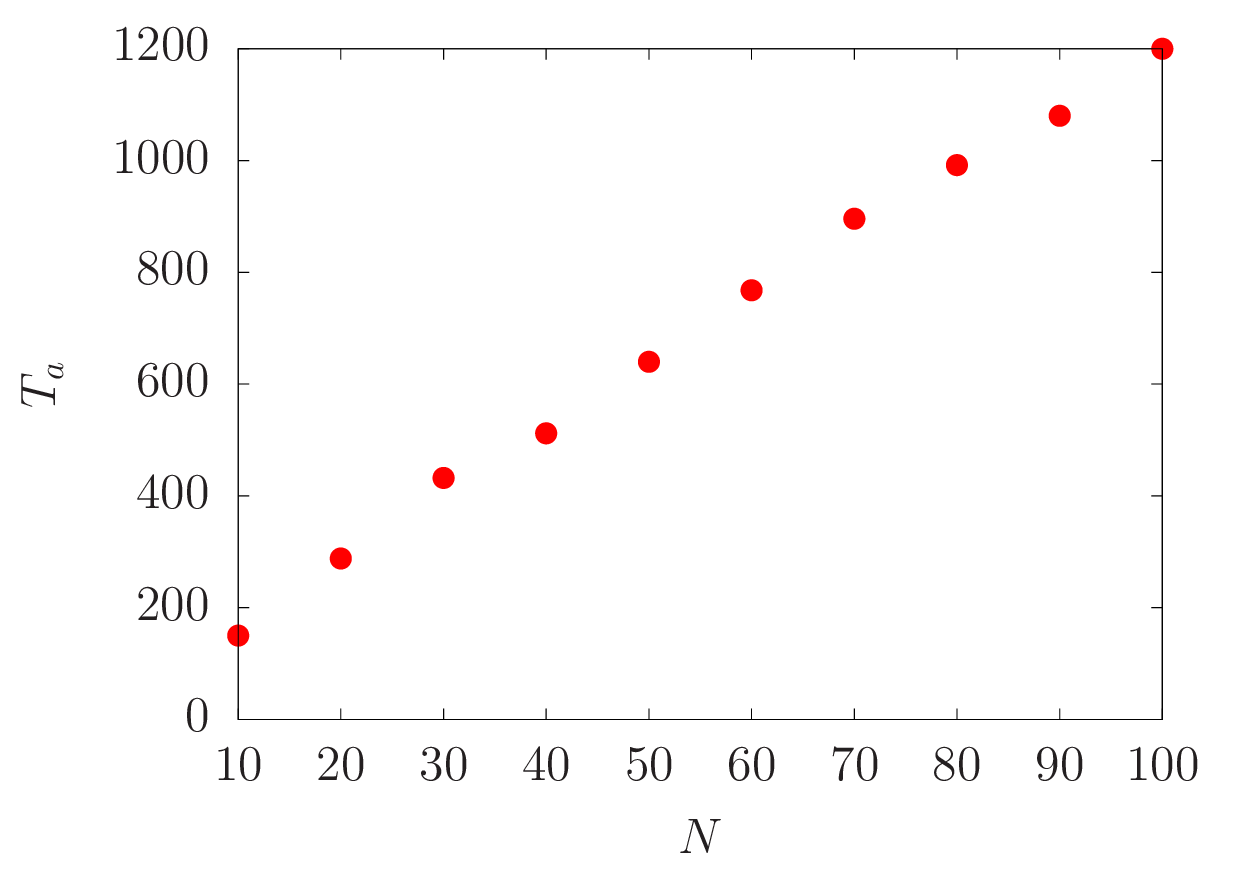}
\caption{\label{Fig.optime}Some examples of locally optimal operation time $T_a$ with respect to the system size $N$. The vertical axis is in units of $(2JN^2)^{-1}$. }
\end{figure}

We also discuss these quantities for other system size, $N=20,30,40,\dots,100$. 
Some examples of locally optimal operation time are plotted with respect to the system size $N$ in Fig.~\ref{Fig.optime}. 
It is given by $T_a\sim(11.6N+60.0)(2JN^2)^{-1}=\mathcal{O}(J^{-1}N^{-1})$. 
Typically, the interaction strength $J$ can be $\mathcal{O}(N^{-1})$~\cite{Dooley2016}, and thus it means that we can set $T_\mathrm{prep(read)}=T_a=\mathcal{O}(N^0)$ for small system size. 
This time duration is much faster than that for state preparation and readout satisfying the adiabatic condition, $T_\mathrm{prep(read)}=T_a=\mathcal{O}(N^{2/3})$, and thus we can use much longer time for sensing or increase the number of measurement.

\begin{figure}
\includegraphics[width=8cm]{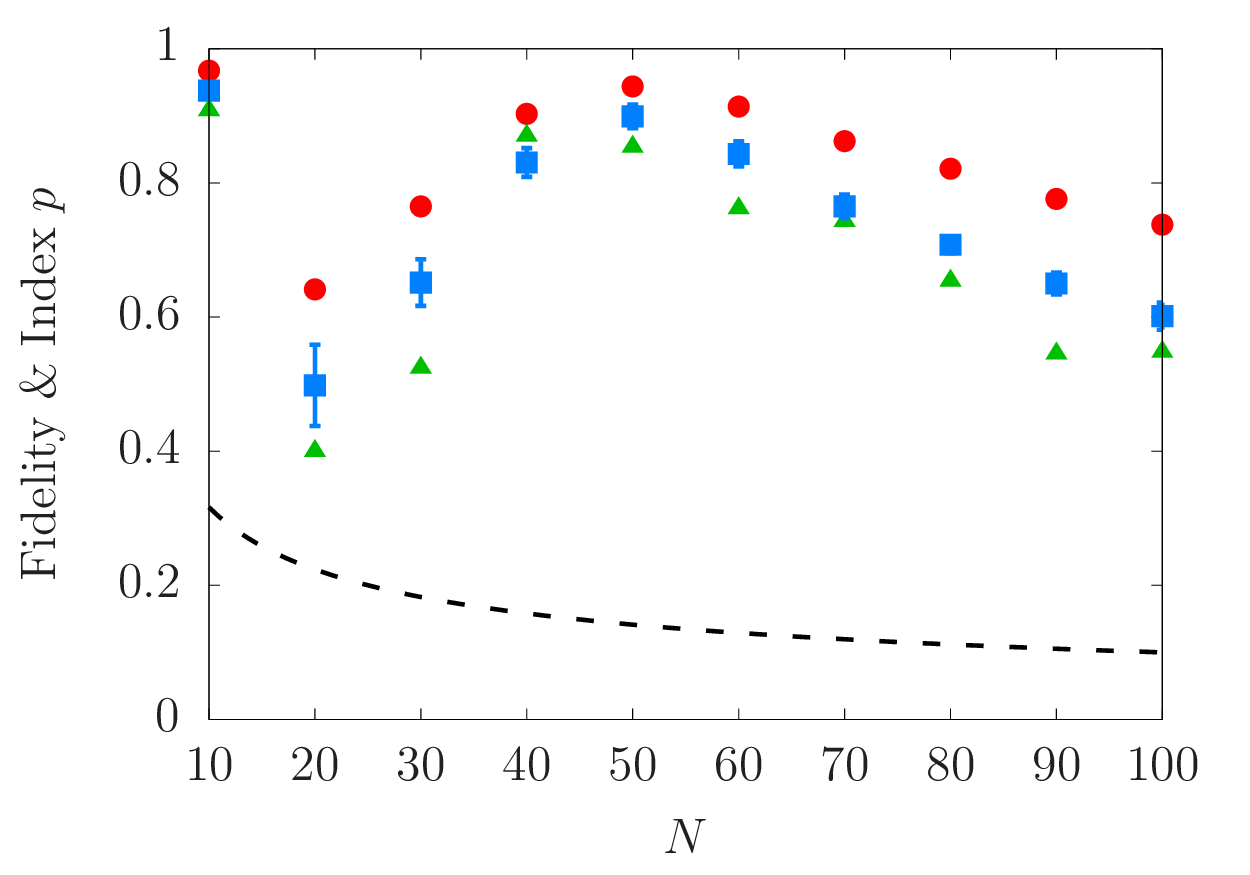}
\caption{\label{Fig.syssize} System size dependence of (red circles) the fidelity of the probe state at time $t=T_a$ to the GHZ state, (green triangles) that of the probe state at time $t=2T_a$ with $T_\mathrm{int}=0$ to the initial state, and (blue squares) the index $p$, which shows how close the uncertainty is to the Heisenberg limit on average. Here we use locally optimal time $T_a$ plotted in Fig.~\ref{Fig.optime}. The error bar in the index $p$ represents standard deviation and the dotted curve represents the SQL. 
}
\end{figure}

By using these locally optimal operation time, we calculate the uncertainty for several $N$ against 
$T_\mathrm{int}$ (see Appendix~\ref{Sec.othersize}). We find that the uncertainty has some dependence on $T_\mathrm{int}$ and it slightly deviates from the Heisenberg limit. 
We express the average uncertainty of estimation as $\delta h^z_\mathrm{est}=1/2pNT_\mathrm{int}$, where $p$ ($0\le p\le1$) is an index denoting how close the uncertainty is to the Heisenberg limit. 
Here, the uncertainty is averaged for $(2JN^2)T_\mathrm{int}=1,3,5,\dots,199$. 
For locally optimal time $T_a$ in Fig.~\ref{Fig.optime}, the fidelity to the GHZ state (red circles), that to the initial state (green triangles), and the index $p$ (blue squares) are calculated and plotted in Fig.~\ref{Fig.syssize}. 
These quantities show complicated behavior against the number of qubits because of nonadiabatic transitions and interference. 
Remarkably, the uncertainty surpasses the SQL. 
Note that the shown performance is not the best; there exists other longer operation time showing better performance. 
If coherent time is long enough, we can choose those operation time.

%
%
\section{\label{Sec.discuss}Summary}

We considered quantum metrology based on symmetry-protected adiabatic transformation. 
In this protocol, parity measurement, which is difficult to be implemented in experiments, is replaced with simple global magnetization measurement by adiabatic transformation of the transverse field. 
Here, we exploited the fact that the parity is a conserved quantity because of the spin-flip symmetry. 
We discussed the effects of the finite transverse field and nonadiabatic transitions as imperfection of adiabatic transformation. 
By taking into account finite time duration for state preparation, sensing, and readout, we also compared performance of the present scheme with the classical scheme in the absence and presence of dephasing.

In this paper, we considered the finite transverse field, finite time duration for state preparation, sensing, and readout, dephasing, and nonadiabatic transitions as possible situations. 
We leave effects of other errors and noises as future work, but we mention some evidence of robustness against various errors and noises. 
Our protocol utilizes the ground state, and thus decay from excited states to lower energy states during entanglement generation is less problematic than conventional dynamical approaches.
In addition, the offset discussed in Sec.~\ref{Sec.noise} makes our protocol robust against measurement imperfection as in the case of the finite transverse field, dephasing, and nonadiabatic transitions. 
Robustness of dynamics against bias, which breaks symmetry-protected conservation laws, during symmetry-protected adiabatic transformation was discussed in Ref.~\cite{Zhuang2020}. 
Robustness of entanglement generation against a loss process, which breaks a symmetry-protected conservation law and confinement in subspace of the Hilbert space, during (super)adiabatic transformation was discussed in Ref.~\cite{Hatomura2019}. 
Symmetry-protected superadiabatic transformation~\cite{Hatomura2018a,Hatomura2019} based on shortcuts to adiabaticity~\cite{Guery-Odelin2019} can also speedup the present protocol and reduce negative effects.


\begin{acknowledgments}
This work was supported by JST PRESTO Grant No.~JPMJPR1919, JST CREST Grant No.~JPMJCR1774, and Leading Initiative for Excellent Young Researchers, MEXT, Japan.
MT is supported by JSPS fellowship (JSPS KAKENHI Grant No. 20J01757).
\end{acknowledgments}

%
%
\appendix

%
%

\section{\label{Sec.meas.Sz}Global magnetization measurement of $\hat{S}_Z$ discussed in Ref.~\cite{Huang2018}}

In Ref.~\cite{Huang2018}, global magnetization measurement of $\hat{S}_Z$ was discussed for the probe state (\ref{Eq.readoutstate}). 
Its measurement outcome is
\begin{equation}
\langle\hat{S}_Z\rangle_{\theta=2h^zT_\mathrm{int}}=\sqrt{\frac{N}{2}}\cos\alpha\sin(2h^zNT_\mathrm{int}),
\end{equation}
and its standard deviation is
\begin{equation}
(\Delta\hat{S}_Z)_{\theta=2h^zT_\mathrm{int}}=\sqrt{\frac{N}{2}}. 
\end{equation}
Therefore, the uncertainty of the estimation is given by
\begin{equation}
\delta h^z_\mathrm{est}=\frac{(\Delta\hat{S}_Z)_{\theta=2h^zT_\mathrm{int}}}{\left|\frac{\partial\langle\hat{S}_Z\rangle_{\theta=2h^zT_\mathrm{int}}}{\partial h^z}\right|\sqrt{M}}=\frac{1}{2N\sqrt{M}T_\mathrm{int}|\cos\alpha\cos(2h^zNT_\mathrm{int})|}, 
\end{equation}
i.e., it satisfies the Heisenberg limit, but cancellation of the relative phase $\alpha$ is necessary and its sensing range is limited due to the factor $\cos(2h^zNT_\mathrm{int})$ even in the ideal situation.

%
%
\section{\label{Sec.bound.derivation}Derivation of the bound (\ref{Eq.uncertainty.bound})}

Equation (\ref{Eq.projection.realistic}) is rewritten as
\begin{equation}
\begin{aligned}
P=&\left|\sum_{n=0}^{N/2}|g_n|^2e^{i\gamma_n}\cos[h^z(N-2n)T_\mathrm{int}]\right|^2 \\
=&\sum_{m,n=0}^{N/2}|g_mg_n|^2e^{i(\gamma_m-\gamma_n)}\cos[h^z(N-2m)T_\mathrm{int}]\cos[h^z(N-2n)T_\mathrm{int}] \\
=&\frac{1}{2}\sum_{m,n=0}^{N/2}|g_mg_n|^2e^{i(\gamma_m-\gamma_n)}\{\cos[2h^z(N-m-n)T_\mathrm{int}]+\cos[2h^z(m-n)T_\mathrm{int}]\}. 
\end{aligned}
\end{equation}
Now we estimate the denominator of Eq.~(\ref{Eq.uncertainty}). 
By using the triangle inequality, we obtain
\begin{equation}
\begin{aligned}
\left|\frac{\partial P}{\partial h^z}\right|=&\left|\sum_{m,n=0}^{N/2}|g_mg_n|^2e^{i(\gamma_m-\gamma_n)}\{(N-m-n)T_\mathrm{int}\sin[2h^z(N-m-n)T_\mathrm{int}]+(m-n)T_\mathrm{int}\sin[2h^z(m-n)T_\mathrm{int}]\}\right| \\
\ge&|g_0|^4NT_\mathrm{int}|\sin(2h^zNT_\mathrm{int})| \\
&-\bigg|\sum_{m,n=0}^{N/2}(1-\delta_{m0}\delta_{n0})|g_mg_n|^2e^{i(\gamma_m-\gamma_n)} \\
&\quad\times\{(N-m-n)T_\mathrm{int}\sin[2h^z(N-m-n)T_\mathrm{int}]+(m-n)T_\mathrm{int}\sin[2h^z(m-n)T_\mathrm{int}]\}\bigg| \\
\ge&|g_0|^4NT_\mathrm{int}|\sin(2h^zNT_\mathrm{int})| \\
&-\sum_{m,n=0}^{N/2}(1-\delta_{m0}\delta_{n0})|g_mg_n|^2|(N-m-n)T_\mathrm{int}\sin[2h^z(N-m-n)T_\mathrm{int}]+(m-n)T_\mathrm{int}\sin[2h^z(m-n)T_\mathrm{int}]| \\
=&|g_0|^4NT_\mathrm{int}|\sin(2h^zNT_\mathrm{int})| \\
&-\left(\sum_{m,n=1}^{N/2}\delta_{mn}+2\sum_{\substack{m,n=0 \\ (m>n)}}^{N/2}\right)|g_mg_n|^2|(N-m-n)T_\mathrm{int}\sin[2h^z(N-m-n)T_\mathrm{int}]+(m-n)T_\mathrm{int}\sin[2h^z(m-n)T_\mathrm{int}]|. 
\end{aligned}
\end{equation}
When the condition $0\le2h^zNT_\mathrm{int}\le\pi/2$ is satisfied, the following inequalities hold: 
\begin{equation}
0\leq (N-m-n)\sin[2h^z(N-m-n)T_\mathrm{int}]\le(N-m)\sin(2h^zNT_\mathrm{int})
\end{equation}
and
\begin{equation}
0\leq (m-n)\sin[2h^z(m-n)T_\mathrm{int}]\le m\sin(2h^zNT_\mathrm{int}) 
\end{equation}
for $0\le m,n\le N/2$ and $m\ge n$. 
Therefore, we obtain
\begin{equation}
\begin{aligned}
&\left(\sum_{m,n=1}^{N/2}\delta_{mn}+2\sum_{\substack{m,n=0 \\ (m>n)}}^{N/2}\right)|g_mg_n|^2|(N-m-n)T_\mathrm{int}\sin[2h^z(N-m-n)T_\mathrm{int}]+(m-n)T_\mathrm{int}\sin[2h^z(m-n)T_\mathrm{int}]| \\
\le&\left(\sum_{m,n=1}^{N/2}\delta_{mn}+2\sum_{\substack{m,n=0 \\ (m>n)}}^{N/2}\right)|g_mg_n|^2N\sin(2h^zNT_\mathrm{int}). 
\end{aligned}
\end{equation}
Moreover, we find
\begin{equation}
\begin{aligned}
\left(\sum_{m,n=1}^{N/2}\delta_{mn}+2\sum_{\substack{m,n=0 \\ (m>n)}}^{N/2}\right)|g_mg_n|^2&=\sum_{m,n=0}^{N/2}(1-\delta_{m0}\delta_{n0})|g_mg_n|^2 \\
&=1-|g_0|^4. 
\end{aligned}
\end{equation}
Finally we obtain
\begin{equation}
\left|\frac{\partial P}{\partial h^z}\right|\ge NT_\mathrm{int}(2|g_0|^4-1)\sin(2h^zNT_\mathrm{int}) 
\end{equation}
for $|g_0|^4>1/2$. 
For $|g_0|^4\le1/2$, we use $|\partial P/\partial h^z|\ge0$ and it results in a trivial bound $\delta h^z_\mathrm{est}\le\infty$. 
Together with a trivial inequality $\sqrt{P(1-P)}\le1/2$, we find a bound for the uncertainty of estimation (\ref{Eq.uncertainty})
\begin{equation}
\delta h^z_\mathrm{est}\le\frac{1}{2N\sqrt{M}T_\mathrm{int}(2|g_0|^4-1)\sin(2h^zNT_\mathrm{int})} 
\end{equation}
for $|g_0|^4\ge1/2$. 
This is the derivation of the bound (\ref{Eq.uncertainty.bound}).

%
%
\section{\label{Sec.ad.time}Adiabatic time scale}

We rewrite the Hamiltonian (\ref{Eq.ham}) as $\hat{\mathcal{H}}=(2JN)\hat{\mathcal{H}}_\mathrm{dl}$, where $\hat{\mathcal{H}}_\mathrm{dl}$ is the dimensionless Hamiltonian $\hat{\mathcal{H}}_\mathrm{dl}=-(1/N)\hat{S}_Z^2-(h^x/JN)\hat{S}_X$. 
In Ref.~\cite{Caneva2008}, linear change of the transverse field $h^x/JN=(h_0^x/JN)(1-2JNt/2JNT_\mathrm{prep})$ was discussed with this Hamiltonian $\hat{\mathcal{H}}_\mathrm{dl}$ and the scaling of the dimensionless residual energy $E_\mathrm{res}/N\approx(h_0^x/JN)^{3/2}(2JNT_\mathrm{prep})^{-3/2}$ was found in the adiabatic time scale, where $E_\mathrm{res}=\langle\Psi|\hat{\mathcal{H}}_\mathrm{dl}|\Psi\rangle+N/4$ with a time-evolved state $|\Psi\rangle$. 
In the adiabatic time scale, the state is approximately given by $|\Psi\rangle=\sqrt{1-\epsilon^2}|\psi_0(0)\rangle+\epsilon|\psi_1(0)\rangle$, and then the residual energy is given by $E_\mathrm{res}=\epsilon^2+\mathcal{O}(\epsilon^2N^{-1})$. 
This result implies $2JNT_\mathrm{prep}\approx\epsilon^{-4/3}(h_0^x/JN)N^{2/3}$ for fixed $\epsilon$. 
In terms of $\bar{C}$ in the main text, we find $\bar{C}\approx\epsilon^{-4/3}/2$. 
For example, when the probability for finding the ground state, $(1-\epsilon^2)$, is $95\%$, it is given by $\bar{C}\approx3.68$, and when it is $99\%$, $\bar{C}\approx10.8$. 
Therefore, we find $\bar{C}=\mathcal{O}(1)$ (or, when we require very high fidelity, $\bar{C}=\mathcal{O}(10)$).

%
%
\section{\label{Sec.finite.time.duration}Conditions for beating the SQL and achieving the Heisenberg limit in the different definition}

In the main text, we show the conditions for beating the minimized SQL (\ref{Eq.SQL.min}), and achieving the minimized Heisenberg limit (\ref{Eq.HL.min2}) under the assumption $T\gg T_\mathrm{prep}+T_\mathrm{int}+T_\mathrm{read}$, where a statistical average is taken into account. 
Here we show that we can obtain the same conditions even if we consider the other minimized Heisenberg limit (\ref{Eq.HL.min}), which is frequently used, but a statistical average is ignored.

In this case, we can set $T_\mathrm{int}=T-2T_a$. 
Then, we find that the condition for beating the SQL, i.e., $\delta h^z_\mathrm{est}<\delta h^z_\mathrm{SQL,min}$, is given by 
\begin{equation}
T_\mathrm{int}>\frac{2T_aN^{-1/2}}{1-N^{-1/2}}=CN^{1/6}+\mathcal{O}(N^{-1/3}). 
\end{equation}
Next, we discuss how the uncertainty is improved and when the Heisenberg limit scaling is achieved. 
The uncertainty of the estimation can be rewritten as $\delta h^z_\mathrm{est}=\delta h^z_\mathrm{SQL,min}/\eta$ and $\delta h^z_\mathrm{est}=\delta h^{z}_\mathrm{HL.,min}/\eta^\prime$, where
\begin{equation}
\eta=\frac{N^{1/2}T_\mathrm{int}}{T_\mathrm{int}+2T_a}
\end{equation}
and
\begin{equation}
\eta^\prime=\frac{T_\mathrm{int}}{T_\mathrm{int}+2T_a},
\end{equation}
respectively. 
We set $T_\mathrm{int}=CN^{1/6+\epsilon}$ with $\epsilon\ge0$. 
We find that 
\begin{equation}
\eta=\frac{N^{\epsilon}}{1+N^{-1/2+\epsilon}}=N^{\epsilon}+\mathcal{O}(N^{-1/2+2\epsilon})
\end{equation}
for $0\le\epsilon<1/2$, and
\begin{equation}
\eta^\prime=\frac{1}{2}
\end{equation}
for $\epsilon=1/2$. 
Therefore, we can beat the SQL and improve the uncertainty by $N^\epsilon$ for $0\le\epsilon<1/2$, and achieve the Heisenberg limit scaling when $\epsilon=1/2$.

%
%
\section{\label{Sec.othersize}Other system size}

\begin{figure}
\includegraphics[width=8cm]{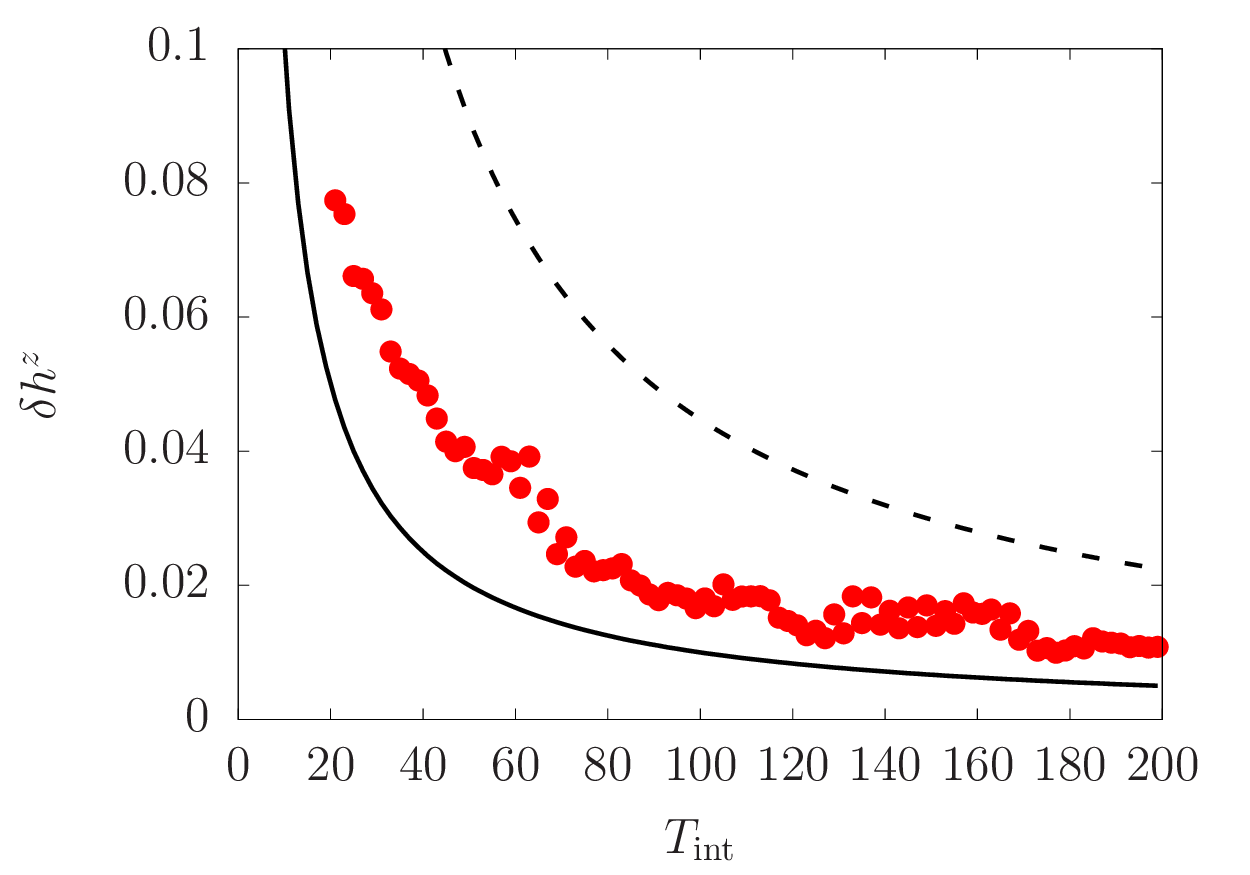}
\includegraphics[width=8cm]{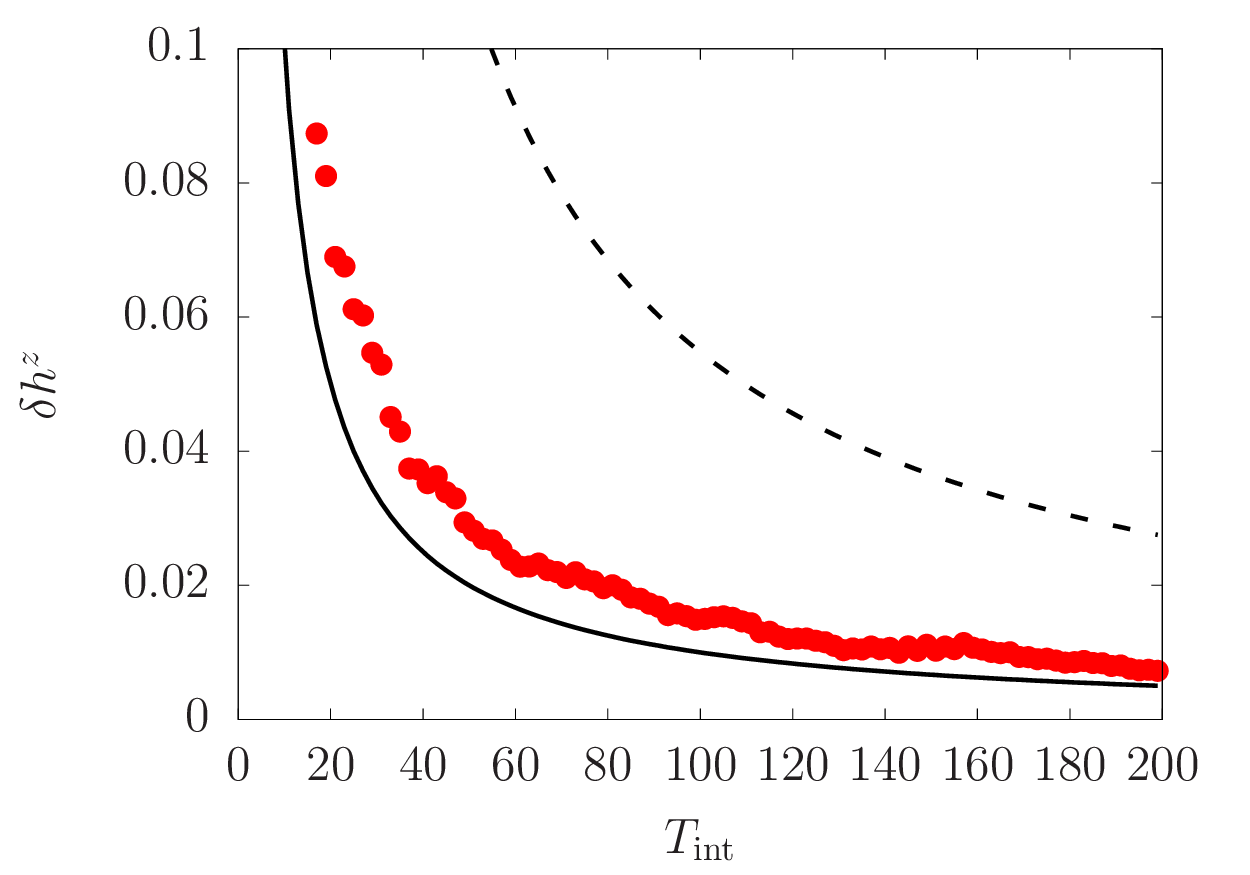}
\includegraphics[width=8cm]{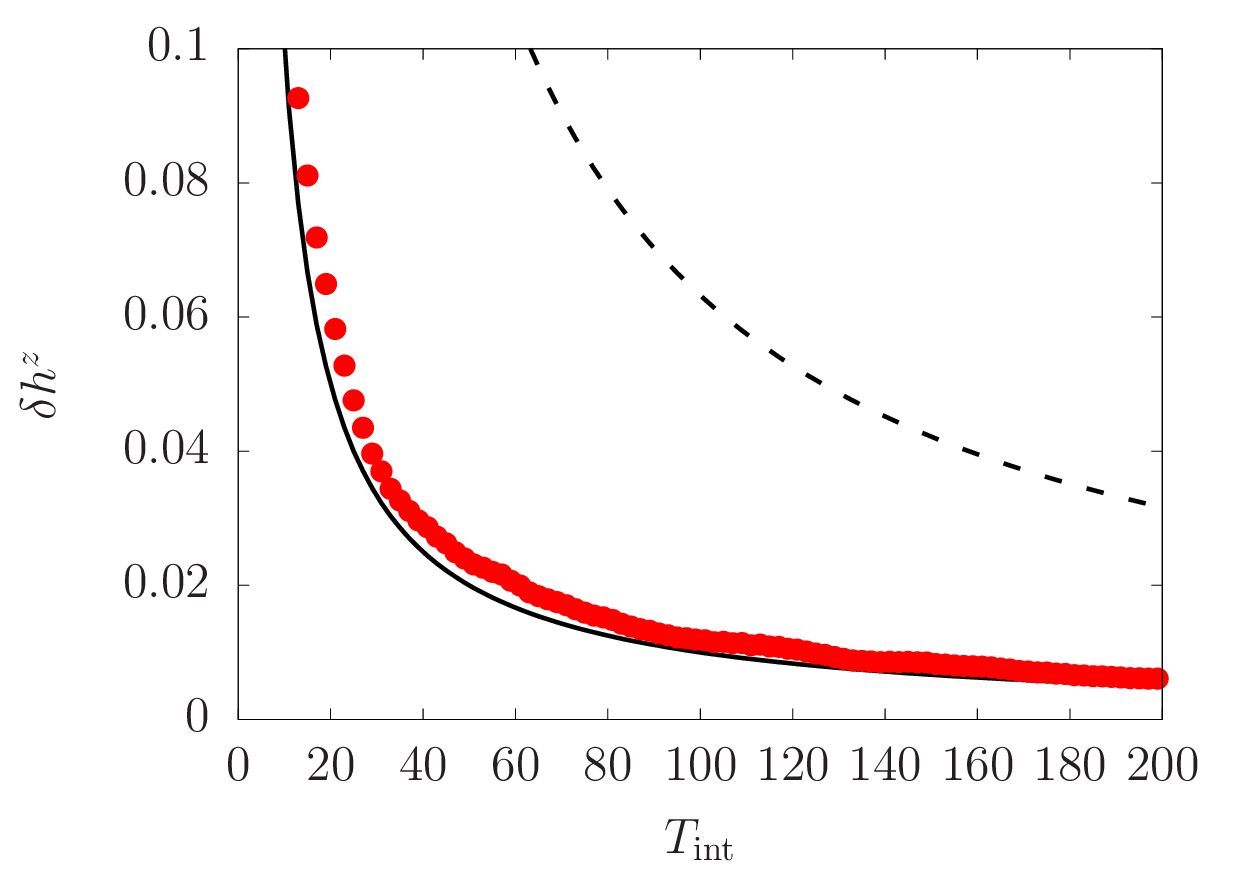}
\includegraphics[width=8cm]{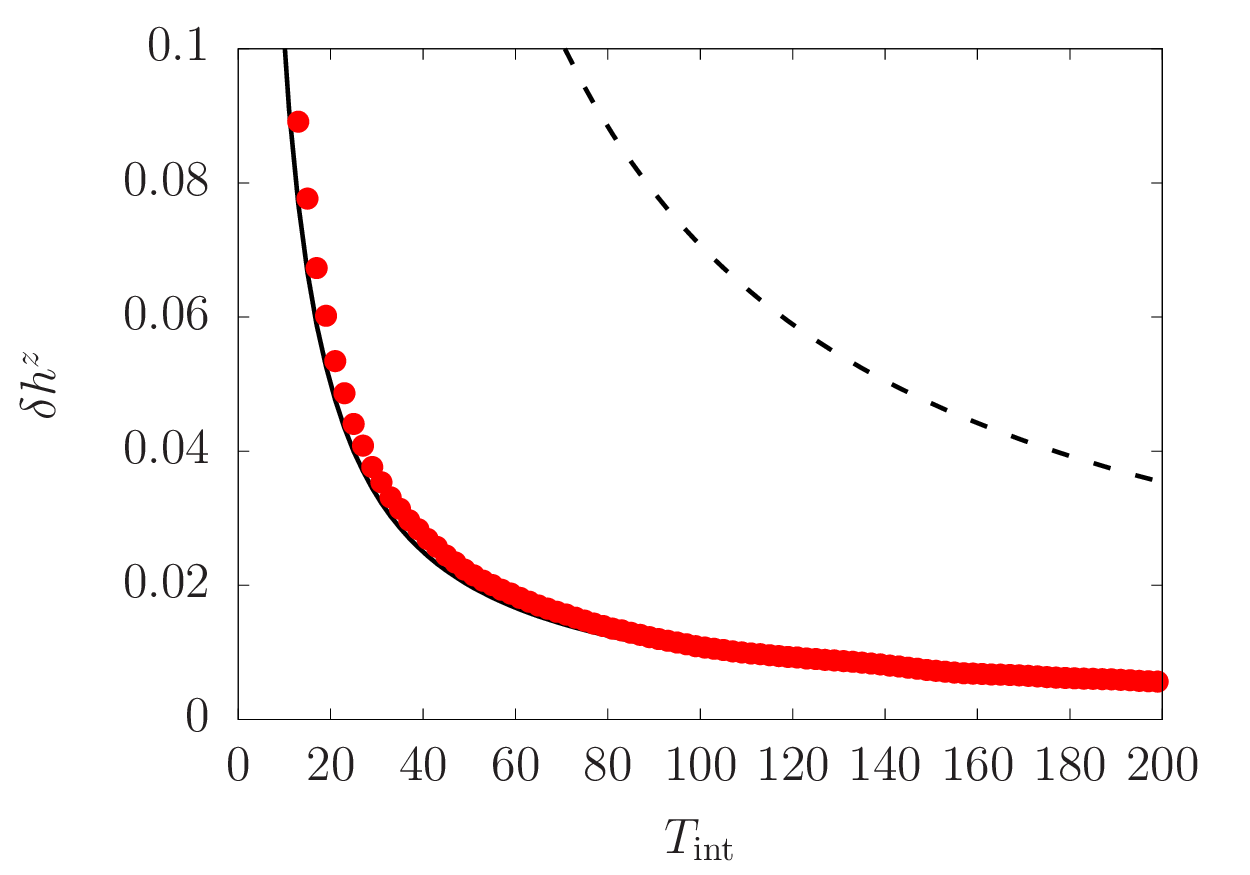}
\caption{\label{Fig.app.othersize}Uncertainty of the estimation for (top left) $N=20$, (top right) $N=30$, (bottom left) $N=40$, and (bottom right) $N=50$. The horizontal axis is the sensing time $T_\mathrm{int}$ in units of $(2JN^2)^{-1}$ and the vertical axis is the uncertainty of estimation $\delta h^z_\mathrm{est}$ in units of $JN$. The solid and dashed curves represent the Heisenberg limit, $\delta h^z_\mathrm{HL}$, and the SQL, $\delta h^z_\mathrm{SQL}$, respectively.}
\end{figure}

Similarly to the case for $N=10$ in Sec.~\ref{Sec.nonad}, we also calculate the uncertainty of the estimation for $N=20$, $30$, $40$, and $50$ against the sensing time $T_\mathrm{int}$ in Fig.~\ref{Fig.app.othersize}, as examples. 
We find that, depending on system size, dependence of the uncertainty on the sensing time $T_\mathrm{int}$ is not negligible.
Therefore, we discuss the uncertainty of the estimation averaged over the sensing time $T_\mathrm{int}$ in the main text.

\bibliography{AQMbib}

\end{document}